\documentclass[conference]{IEEEtran}
\IEEEoverridecommandlockouts

\usepackage{cite}
\usepackage{url}
\usepackage{amsmath,amsfonts} 
\usepackage{float}
\usepackage{graphicx}
\usepackage{algorithm}
\usepackage{algpseudocode}
\usepackage{siunitx}
\usepackage[labelfont=bf]{caption}
\usepackage{subcaption}
\usepackage{tikz}
\usepackage{pgfplots}
\usepackage{epstopdf}
\usepackage{multirow}
\usepackage[font=scriptsize]{caption}
\usepackage{soul}
\usepackage{color, colortbl}
\usepackage{array}
\usepackage[T1]{fontenc}
\usepackage[utf8]{inputenc}
\usepackage{mathptmx}
\usepackage{listings}
\usepackage{multirow}

\usepackage{enumitem}

\def\BibTeX{{\rm B\kern-.05em{\sc i\kern-.025em b}\kern-.08em
    T\kern-.1667em\lower.7ex\hbox{E}\kern-.125emX}}

\begin{document}

\title{EDIMA: Early Detection of IoT Malware Network Activity Using Machine Learning Techniques}
			
\author{\IEEEauthorblockN{Ayush Kumar and Teng Joon Lim}
\IEEEauthorblockA{Department of Electrical and Computer Engineering\\
\textit{National University of Singapore}\\
Singapore\\
ayush.kumar@u.nus.edu, eleltj@nus.edu.sg}
}

\IEEEoverridecommandlockouts
\IEEEpubid{\makebox[\columnwidth]{978-1-5386-4980-0/19/\$31.00 \copyright2019 IEEE} \hspace{\columnsep}\makebox[\columnwidth]{ }}

\maketitle

\begin{abstract}
The widespread adoption of Internet of Things has led to many security issues. Post the Mirai-based DDoS attack in 2016 which compromised IoT devices, a host of new malware using Mirai's leaked source code and targeting IoT devices have cropped up, e.g. Satori, Reaper, Amnesia, Masuta etc. These malware exploit software vulnerabilities to infect IoT devices instead of open TELNET ports (like Mirai) making them more difficult to block using existing solutions such as firewalls. In this research, we present EDIMA, a distributed modular solution which can be used towards the detection of IoT malware network activity in large-scale networks (e.g. ISP, enterprise networks) during the scanning/infecting phase rather than during an attack. EDIMA employs machine learning algorithms for edge devices' traffic classification, a packet traffic feature vector database, a policy module and an optional packet sub-sampling module. We evaluate the classification performance of EDIMA through testbed experiments and present the results obtained. 
\end{abstract}


\section{Introduction}
The Internet of Things (IoT)\cite{iotsurvey} is a network of sensing devices with limited resources and capable of wired/wireless communications with cloud services. IoT devices are being increasingly targeted by attackers using malware as they are easier to infect than conventional computers. This is due to several reasons\cite{iotsecsurvey1} such as presence of legacy devices with no security updates, low priority given to security within the development cycle, weak login credentials, etc. 

In a widely publicized attack, the IoT malware \textit{Mirai} was used to propagate the biggest DDoS (Distributed Denial-of-Service) attack on record on October 21, 2016. The attack targeted the Dyn DNS (Domain Name Service) servers \cite{miraiattack} and generated an attack throughput of the order of 1.2 Tbps. It disabled major internet services such as Amazon, Twitter and Netflix. The attackers had infected IoT devices such as IP cameras and DVR recorders with Mirai, thereby creating an army of bots (botnet) to take part in the DDoS attack.

The source code for Mirai was leaked in 2017 and since then there has been a proliferation of IoT malware. Script ``kiddies'' as well as professional blackhat/greyhat hackers have used the leaked source code to build their own IoT malware. These malware are usually variants of Mirai using a similar brute force technique of scanning random IP addresses for open TELNET ports and attempting to login using a built-in dictionary of commonly used credentials (Remaiten, Hajime), or more sophisticated malware that exploit software vulnerabilities to execute remote command injections on vulnerable devices (Reaper, Satori, Masuta, Linux.Darlloz, Amnesia etc.). Even though TELNET port scanning can be countered by deploying firewalls (at the user access gateway) which block incoming/outgoing TELNET traffic, malware exploiting software vulnerabilities involving application protocols such as HTTP, SOAP, PHP etc. are more difficult to block using firewalls because those application protocols form a part of legitimate traffic as well.
 
Bots compromised by Mirai or similar IoT malware can be used for DDoS attacks, phishing and spamming \cite{phishspam}. These attacks can cause network downtime for long periods which may lead to financial loss to network companies, and leak users' confidential data. Bitdefender mentioned in its blog in September 2017\cite{bitdef} that researchers had estimated at least 100,000 devices infected by Mirai or similar malware revealed daily through TELNET scanning telemetry data. In an October 2017 article\cite{arbornetw}, Arbor researchers estimated that the actual size of the Reaper botnet fluctuated between 10,000-20,000 bots but warned that this number could change at any time with an additional 2 million devices having been identified by botnet scanners as potential Reaper bots. A Kaspersky lab report\cite{kasperskyrep} released in September 2018 says that 121,588 IoT malware samples were identified in the first half of 2018 which was three times the number of IoT malware samples in the whole of 2017. 

Further, many of the infected devices are expected to remain infected for a long time. Therefore, there is a substantial motivation for detecting these IoT bots and taking appropriate action against them so that they are unable to cause any further damage. As pointed out in \cite{trillionflaws}, attempting to ensure that all IoT devices are secure-by-construction is futile and it is practically unfeasible to deploy traditional host-based detection and prevention mechanisms such as antivirus, firewalls for IoT devices. Therefore, it becomes imperative that the security mechanisms for the IoT ecosystem are designed to be network-based rather than host-based.  

In this research, we propose a solution towards detecting the network activity of IoT malware in large-scale networks such as enterprise and ISP (Internet Service Provider) networks. Our proposed solution consists of machine learning (ML) algorithms running at the user access gateway which detect malware activity based on their scanning traffic patterns, a database that stores the malware scanning traffic patterns and can be used to retrieve or update those patterns, and a policy module which decides the further course of action after gateway traffic has been classified as malicious. It also includes an optional packet sub-sampling module which can be deployed for example, in case of enterprises where a number of IoT devices ($\approx$ 10-100) are connected to a single access gateway. The bot detection solution can be deployed both on physical access gateways supplied by the ISP companies or as NFV (Network Function Virtualization) functions at the customer premises/enterprise in a SDN-NFV based network architecture, where SDN stands for Software-Defined Networking.

Bots scanning for and infecting vulnerable devices are targeted in particular by our solution. This is because the scanning and propagation phase of the botnet life-cycle stretches over many months and we can detect and isolate the bots before they can participate in an actual attack such as DDoS. If the DDoS attack has already occurred (due to a botnet), detecting the attack itself is not that difficult and there are already existing methods both in literature and industry to defend against such attacks. Once the IoT bots are detected, the network operators can take suitable countermeasures such as blocking the traffic originating from IoT bots and notifying the local network administrators.
The major contributions of this paper are listed below:
\begin{enumerate}
\item We have categorized most of the current IoT malware into a few categories to help identify similar malware and simplify the task of designing detection methods for them.
\item We have analyzed the traffic patterns for IoT malware from each category through testbed experiments and packet capture utilities. 
\item We have proposed a modular solution towards detection of IoT malware activity by using ML techniques with the above traffic patterns. 
\end{enumerate}


\section{Related Work}
\label{literature}
There are several works in the literature on detecting PC-based botnets using their CnC (Command-and-control) server communication features. 
Bothunter\cite{bothunter} builds a \textit{bot infection dialog model} based on which three bot-specific sensors are constructed and correlation is performed between inbound intrusion/scan alarms and the infection dialog model to generate a consolidated report. Spatio-temporal similarities between bots in a botnet in terms of bot-CnC coordinated activities are captured from network traffic and leveraged towards botnet detection in a local area network in Botsniffer\cite{botsniffer}. In BotMiner\cite{botminer}, the authors have proposed a botnet detection system which clusters similar CnC communication traffic and similar malicious activity traffic, and uses cross cluster correlation to detect bots in a monitored network.

There has also been some research on intrusion detection and anomaly detection systems for IoT. A whitelist-based intrusion detection system for IoT devices (Heimdall) has been presented in \cite{heimdall}. 
The authors in \cite{tldrtc} propose an intrusion detection model for IoT backbone networks leveraging two-layer dimension reduction and two-tier classification techniques to detect U2R (User-to-Root) and R2L (Remote-to-Local) attacks. 

Of late, there has been an interest in IoT botnet and attack detection in the research community resulting in a number of papers addressing these problems. In \cite{nbaiot}, deep-autoencoders based anomaly detection has been used to detect attacks launched from IoT botnets. 
A few works have focused on building normal communication profiles for IoT devices which are not expected to deviate much over a long period of time. DEFT\cite{deft} has used ML algorithms at SDN controllers and access gateways to build normal device traffic fingerprints while \cite{mudprofile} proposes a tool to automatically generate MUD (Manufacturer Usage Description) profiles for a number of consumer IoT devices. In DIoT \cite{diot}, the authors have proposed a method to classify typically used IoT devices into various device types and build their normal traffic profiles so that a deviation from those profiles is flagged as anomalous traffic.

Our work addresses a few important gaps in the literature when it comes to distinguishing between legitimate and botnet IoT traffic. First, the works on detecting botnets using their CnC communication features \cite{ircbotnet, bothunter,botsniffer,botminer} are designed for PC-based botnets rather than IoT botnets which are the focus of our work.
Second, we do not aim to detect botnets (networks of bots) but instead, network activity generated by individual bots. IoT botnets tend to consist of hundreds of thousands to millions of devices spread over vast geographies, hence, it is impractical to detect a whole network of IoT bots. Therefore, we do not require computationally expensive clustering algorithms as used in \cite{botsniffer,botminer}. 

Third, unlike \cite{nbaiot,diot}, we aim to detect IoT malware activity much before the actual attack, during the scanning/infection phase. Finally, instead of fingerprinting the normal traffic of IoT devices \cite{deft,diot} and using those fingerprints towards anomaly detection, we detect the malware-induced scanning packet traffic generated by infected IoT devices. This is because the former approach suffers from limitations such as possibility of misclassification of an infected device as a legitimate device type, testing against only simple malware e.g. Mirai which may result in failure to detect other, more sophisticated malware, etc. The latter approach is not free from limitations as well, since it is not resilient against new undiscovered malware whose scanning traffic features have not been updated in the database. We advocate for a combined approach consisting of both IoT device fingerprinting/anomaly detection and IoT malware scanning traffic detection.

\section{EDIMA Architecture}
\begin{figure}[h]
\centering
\includegraphics[scale=0.25]{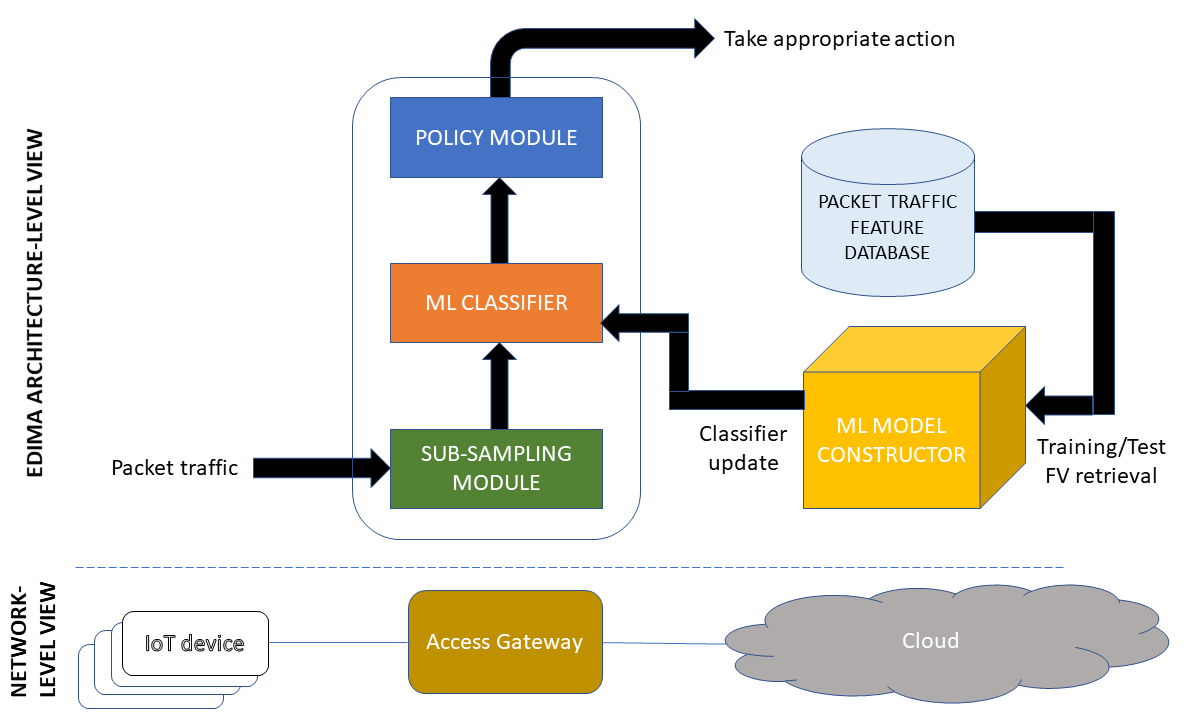}
\caption{EDIMA Architecture}
\label{EDIMA-arch}
\end{figure}

Our proposed solution towards detecting the scanning packet traffic generated by IoT malware through the use of ML algorithms is called EDIMA (Early Detection of IoT Malware Network Activity) and is shown in Fig.\ref{EDIMA-arch}. It is designed to have a modular architecture with of five different modules:
\begin{enumerate}
\item \textbf{ML Classifier}:
The ML classifier runs on the access gateway connected to IoT devices at customer premises or enterprise. It collects the incoming traffic samples, extracts the feature vectors for those samples and classifies them based on the ML model trained by \textit{ML model constructor}. More details about the ML classifier are given in Section \ref{classifierdetails}.

\item \textbf{ML Model Constructor}:
The ML model for classifying access gateway traffic is trained by ML model constructor using the feature vectors and class labels retrieved from \textit{Packet traffic feature database} as inputs to a supervised classification algorithm such as Naive Bayes (NB), Decision Trees (DT), Support Vector Machines (SVM) etc. The model is then sent to the \textit{ML classifier}. Whenever a new malware is discovered, the ML model has to be re-trained and compared with the existing ML model for classification performance. If there is no significant improvement in performance, the existing ML model continues to be used, otherwise the re-trained ML model is updated to the \textit{ML classifier} module.

\item \textbf{Packet Traffic Feature Database}:
The database stores a list of feature vectors extracted from traffic samples collected from access gateways connected to IoT devices infected with known IoT malware as well as gateways connected to uninfected devices. The database is updated frequently for newly discovered malware. The feature vectors and corrresponding class labels are retrieved by the \textit{ML model constructor} for training ML classifier for the first time and also for re-training the classifier whenever a new malware is discovered. We envisage a community of security researchers, industry personnel and users who will collect traffic data for IoT malware through honeypots, consumer access gateways etc. The feature vectors extracted from the raw traffic data samples and the class labels assigned to those samples will be updated to the online feature database.
 
\item \textbf{Policy Module}:
The policy module consists of a list of policies defined by network administrator which decide the course of actions to be taken once the traffic from an access gateway has been classified as malicious by the ML classifier module. For instance, the network administrator can block the entire traffic originating from bots and bring them back online only after it is confirmed that the malware has been removed from those IoT devices. 

\item \textbf{Sub-sampling Module} (optional):
For premises having thousands of IoT devices such as enterprises, industries etc. we also propose an optional sub-sampling module as introduced in \cite{ayushficc}. This module samples the packet traffic from IoT devices both along time as well across the devices and presents them as input to the ML classifier module. The sub-sampling module would help reduce the computational overhead for ML classifier module by forwarding only a fraction of the incoming IoT packet traffic.

\end{enumerate}

\section{Extraction of IoT Malware Traffic Features}
\subsection{Malware Categorization}
We have categorized known IoT malware into three categories based on type of vulnerability that they target: TELNET, HTTP POST and HTTP GET. TELNET is an application-layer protocol used for bidirectional byte-oriented communication. Typically, a user with a terminal and running a TELNET client program, accesses a remote host running a TELNET server by requesting a connection to the remote host and logging in by providing its credentials. HTTP GET and POST are methods based on HTTP (HyperText Transfer Protocol) application-layer protocol which are used to request data from and send data to server resources respectively. For example, HTTP GET is commonly used for requesting web pages from remote web servers through a browser. 
We have presented the malware categories, various malware belonging to those categories and brief descriptions of their operation in Table \ref{table-malware}.
\begin{table}
	\centering
	\vspace*{0.5cm}
    \begin{tabular}{ | l | l | p{4cm} | }
    \hline
    \textbf{Category} & \textbf{Malware} & \textbf{Description} \\ \hline
    \multirow{6}{*}{TELNET} & Mirai & Sends SYN packets to probe open TELNET ports at random IP addresses. If successful, it tries to login using list of default credentials\cite{mirai}. \\
 	& Hajime & Same propagation mechanism as Mirai, but no CnC server. Instead, it is built on a P2P network. Purpose seems to be to improve security of IoT devices\cite{hajime}.  \\
 	& Remaiten & Same propagation mechanism as Mirai. Downloads binary specific to targeted platform. Uses IRC protocol for CnC server communication\cite{remaiten}.\\
 	& Linux.Wifatch & Same propagation mechanism as Mirai. Apparently, it tries to secure IoT devices from other malware\cite{linuxwifatch}.  \\
 	& Brickerbot & Rewrites the device firmware, rendering the device permanently inoperable\cite{brickerbot}. \\ \hline	 	
 	    \multirow{4}{*}{HTTP POST} & Satori & Sends NewInternalClient request through miniigd SOAP service (REALTEK SDK) or sends malicious packets to port 37215 (Huwaei home gateway)\cite{satori}. \\
 	& Masuta & Forms SOAP request which bypasses authentication and causes arbitrary code execution\cite{masuta}.  \\
 	& Linux.Darlloz & Sends HTTP POST requests by using  PHP 'php-cgi' Information Disclosure Vulnerability to download the worm from a malicious server on an unpatched device\cite{linuxdarlloz}.  \\ 	
 	& Reaper & Scans first on a list of TCP Ports to fingerprint devices, then second wave of scans on TCP ports running web services such as 80, 8080\dots, sends HTTP POST request for command injection\cite{reaper}.\\ \hline
 	\multirow{2}{*}{HTTP GET} & Reaper & Scanning behavior similar as above, sends HTTP request for remote command execution, usually through CGI or PHP.\\
 	& Amnesia & Makes simple HTTP requests, searches for a special string “Cross Web Server” in the HTTP response from target. If successful, sends four more HTTP requests which contain exploit payloads of four different shell commands\cite{amnesia}.  \\ \hline
    \end{tabular}
    \caption{IoT Malware Categories}
    \label{table-malware}
\end{table}

\subsection{ML Classification}
\label{classifierdetails}
The classification is performed on IoT access gateway-level traffic rather than device-level traffic as working on aggregate traffic is faster and reduces the memory space required. We define two classes of gateway-level  traffic : \textit{benign} and \textit{malicious}. Benign traffic refers to the gateway traffic with no malware-induced scanning packets while malicious traffic refers to gateway traffic that includes malware-induced scanning packets from one of the three malware categories.  For classification of gateway traffic, we have to first generate training data samples consisting of packet captures belonging to those classes. Benign traffic is not difficult to generate since it involves the normal operation of uninfected devices. However, malicious traffic would contain both benign traffic as well as scanning/infection packets generated by malware. To keep things simple, we chose to collect the gateway traffic statically in fixed session intervals. Further, we apply the classification algorithm on these traffic sessions rather than individual packets because per-packet classification is computationally much more costly and doesn't yield any significant benefits. The 

The steps for gateway-level traffic classification are given below:
\begin{enumerate}
\item Filter each traffic session to include only TCP packets with SYN flag activated and destination port numbers belonging to a target list.
\item Extract the feature vectors for each traffic session.
\item Retrieve the trained classifier from ML model constructor and apply it on the extracted feature vectors to classify the corresponding sessions.  
\end{enumerate} 
The target list of destination port numbers is made on the basis of information obtained from public malware exploits. For example, in 'TELNET' category, target destination port numbers are 23 and 2323. In 'HTTP POST' category, target destination port numbers are 37215, 80, 20736, 36895 etc. In 'HTTP GET' category, target destination port number is always 80.


In this work, we use a total of 4 features for ML model training and traffic classification:
\begin{enumerate}
\item Number of unique destination IP addresses
\item Number of packets per destination IP address (maximum, minimum, mean)
\end{enumerate}
The motivation behind selecting the first feature is that the malware generate random IP addresses and send malicious requests to them. Hence, the number of unique destination IP addresses in case of malware-induced scannning traffic will be far more than benign traffic. The second feature set seeks to exploit the fact that malware typically do not send multiple malicious packets to the same IP address (only a single packet is sent in most cases), possibly to cover as many devices as possible during the scanning/propagation phase.

One may argue that the malware author/attacker can adopt a less aggressive scanning strategy to avoid detection. The attacker will incur a cost though, in terms of the malware performance, resulting in fewer infected devices in a fixed time period. We plan to investigate this malware performance-scanning behavior trade off by formulating an optimization problem in the future. For now, the duration of traffic sessions collected for training/classification can be increased to counter any decrease in scanning rates by the attacker.

\section{Performance Evaluation}
\label{eval}
\subsection{Testbed Description}
\begin{figure}[h]
\centering
\includegraphics[scale=0.3]{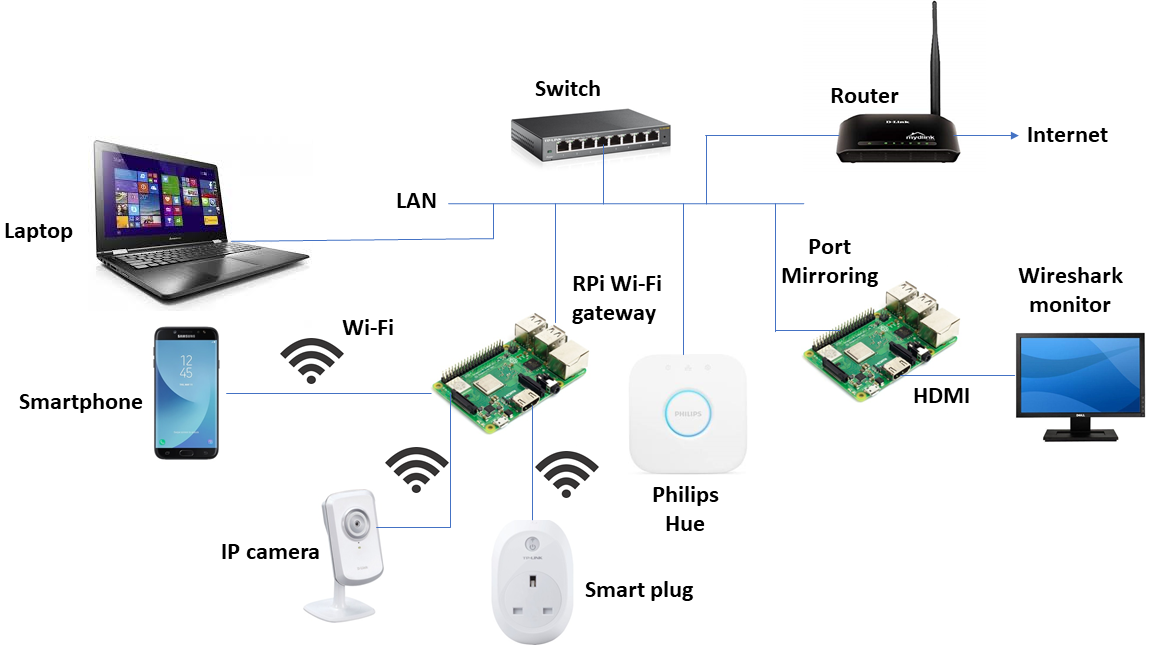}
\caption{Testbed used to collect packet traffic for ML training}
\label{EDIMA-testbed}
\end{figure}

We built a testbed with IoT devices, a laptop PC, Android smartphone and a wireless access gateway to collect ingress/egress traffic at the gateway which would form a part of the training data used to train the ML algorithms to be deployed in the \textit{ML Classifier} module. The IoT devices were: Philips Hue bridge, D-Link DCS-930L Wi-Fi IP camera and TP-Link HS110 Smart Wi-Fi Plug. The laptop PC has an Intel Core i3-5020U 2.2 GHz processor with 4GB RAM and runs Windows 10 OS. Network applications such as web browser (accessing web pages, video streaming sites e.g. YouTube), email client, WiFi camera online platform etc. were run on the the laptop PC by a user. The Android smartphone has Cortex-A53 Octa-core 1.6 GHz processor with 3GB RAM and runs Android 8.0 OS. Again, the same user ran applications such as web browser, social media (Facebook/Twitter/LinkedIn), chat (WhatsApp), Wi-Fi plug app, Hue app etc. on the smartphone which also ran a few other network applications in the background. The wireless access gateway was a D-Link DIR-600 router with an Atheros AR7240 350 MHz network processor, Atheros AR9285 network adapter, 32MB RAM, 4MB flash supporting 1EEE 802.11b/g/n Wi-Fi standards.
The testbed is shown in Fig. \ref{EDIMA-testbed}. We used a TP-Link TL-SG108E Gigabit Ethernet switch with port-mirroring feature to mirror the traffic from all of the above devices (IoT, laptop, smartphone) to a Raspberry Pi 3B+ Ethernet port and monitor the cumulative traffic. 

\subsection{Evaluation Methodology}
As we can't use real malware due to legal and ethical considerations, we wrote scripts to simulate the generation of malicious packets based on publicly available exploits \cite{exploitdb} for the vulnerabilities exploited by those malware. The script generates random IP addresses and sends malicious requests to them in order to execute remote command injection attacks. The injected commands were non-malicious (for ex. \textit{ls -l}, \textit{uname -a}), thus causing no actual harm to any device in the network even if it was vulnerable. The scanning/infection rates in our scripts were designed keeping in the mind the scanning/infection behavior reported online and the Mirai source code which is the basis for most of the current IoT malware. We selected one malware per category for our performance evaluation since the malware in each category have similar scanning/infection behavior.

A total of 60 traffic sessions of 15 minutes duration each were collected for both benign and malicious classes through our testbed. The traffic sessions collected for each case were divided into two sets: \textit{training} and \textit{test} data using a 70:30 split. For the training data, the class labels were assigned to each feature vector extracted from the traffic sessions included in the training data.

\subsection{Results}
\begin{figure}
	\centering
	\begin{subfigure}[b]{0.2\textwidth}
		\includegraphics[width=\textwidth]{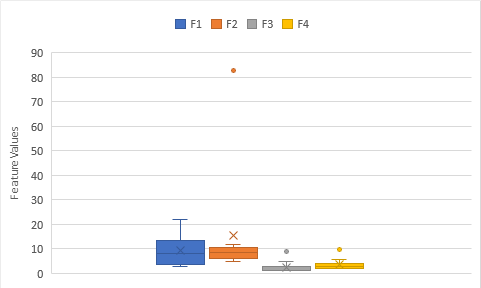}
		\caption{\textit{Benign traffic} training data}
		\label{Benign-TELNET-box-plot}
	\end{subfigure} 
	\begin{subfigure}[b]{0.2\textwidth}
		\includegraphics[width=\textwidth]{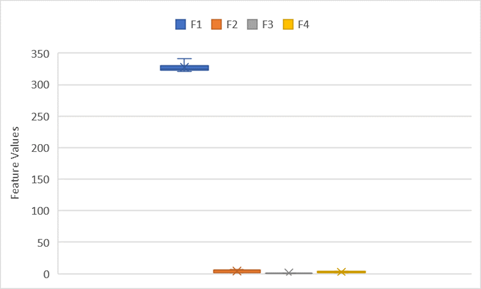}
		\caption{\textit{Malicious TELNET traffic} training data}
		\label{Malicious-TELNET-box-plot}
	\end{subfigure} 
	\begin{subfigure}[b]{0.2\textwidth}
		\includegraphics[width=\textwidth]{Benign-POST-box-plot}
		\caption{\textit{Benign traffic} training data}
		\label{Benign-TELNET-box-plot}
	\end{subfigure} 
	\begin{subfigure}[b]{0.2\textwidth}
		\includegraphics[width=\textwidth]{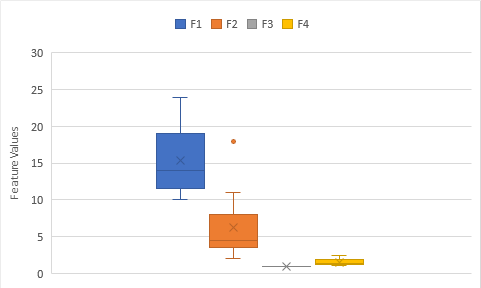}
		\caption{\textit{Malicious POST traffic} training data}
		\label{Malicious-POST-box-plot}
	\end{subfigure} 
	\begin{subfigure}[b]{0.2\textwidth}
		\includegraphics[width=\textwidth]{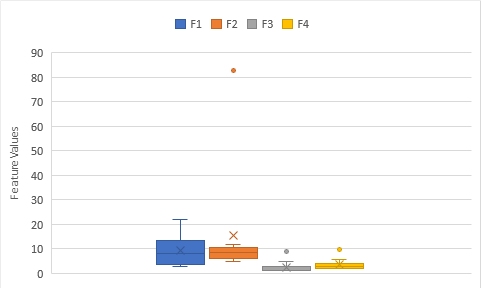}
		\caption{\textit{Benign traffic} training data}
		\label{Benign-TELNET-box-plot}
	\end{subfigure} 
	\begin{subfigure}[b]{0.2\textwidth}
		\includegraphics[width=\textwidth]{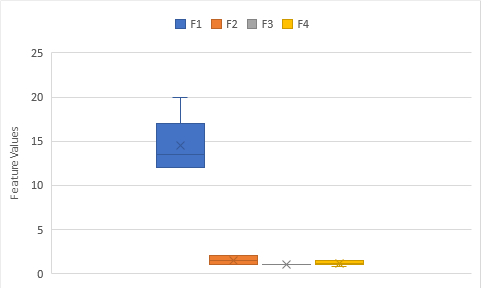}
		\caption{\textit{Malicious GET traffic} training data}
		\label{Malicious-GET-box-plot}
	\end{subfigure}
	\caption{Distribution of feature vector values}
	\label{feature-vec-dist}
\end{figure}

The distributions of the feature values for \textit{benign} and \textit{malicious} training data where the malware belongs to TELNET, HTTP POST and HTTP GET categories are shown in Fig. \ref{feature-vec-dist} using box plots. The distribution plots for feature \textit{F1} under benign and malicious conditions where the malware belongs to TELNET category, are quite visibly distinct, though for the other features \textit{Feature2, Feature3, Feature4}, the plots are not that easily distinguishable. Similarly, the distributions of \textit{HTTP POST} and \textit{GET} packet traffic features under benign and malicious conditions are not completely distinguishable. If there is a significant difference in the distribution of a feature under benign and malicious conditions, that difference can be leveraged by the trained ML model to distinguish between benign and malicious traffic with reasonable detection accuracy. However, if the feature distributions under the two conditions are not easily distinguishable, it may impair the detection performance.

The \textit{scikit-learn} ML algorithms library\cite{scikitlearn} was used for training and classification purposes. We trained Gaussian Naive Bayes, k-NN (k-Nearest Neighbor) and Random Forest algorithms with our training data and evaluated the trained ML models with test data for all three malware categories. The classification accuracy, precision, recall and F-1 scores obtained for the above three classification algorithms are shown in Table \ref{class-scores}. 

The \textit{classification accuracy} refers to the fraction of the total number of input samples whose labels are correctly predicted by a classifier. The \textit{precision} is the ratio $\frac{TP}{TP+FP}$, where $TP$ is the number of true positives and $FP$ is the number of false positives. It represents the ability of a classifier to avoid labeling samples that are negative as positive. The \textit{recall} is the ratio $\frac{TP}{TP+FN}$, where $TP$ is the number of true positives and $FN$ is the number of false negatives. It represents the ability of a classifier to avoid labeling samples that are positive as negative. The \textit{F1 score} is the harmonic mean of precision and recall, expressed as $2\times\frac{{precision}\times{recall}}{precision+recall}$. It represents balance between precision and recall offered by a classifier. The scores in Table \ref{class-scores} show that  the k-NN classifier performs the best followed by Random Forest classifier and Gaussian Naive Bayes classifier.
\begin{table}
	\centering
    \begin{tabular}{ | l | l | l | l | l | }
    \hline
    \textbf{Classifier} & \textbf{Accuracy} & \textbf{Precision} & \textbf{Recall} & \textbf{F1 Score} \\ \hline
    Random Forest & 88.8\% & 0.86 & 1 & 0.92 \\ \hline
    k-NN & 94.44\% & 0.92 & 1 & 0.96 \\ \hline
    Gaussian Naive Bayes & 77.78\% & 0.75 & 1 & 0.86 \\ \hline
    \end{tabular}
    \caption{Accuracy, Precision, Recall and F1 scores for various classifiers }
    \label{class-scores}
\end{table}



\section{Conclusion}
In this paper, we proposed EDIMA, a modular solution for early detection of network activity originating from IoT malware using ML classification techniques. Existing IoT malware were distributed among multiple categories based on their targeted software vulnerabilities. Later, steps for the ML classifier operation and the features used for classification were listed. A testbed consisting of PC, smartphone and IoT devices connected to an access gateway was used to evaluate the classification performance of EDIMA. Using packet traffic captures at access gateway-level, feature vectors were extracted with class labels (\textit{benign} or \textit{malicious}) assigned to them. Subsequently, we depicted the distribution of benign and malicious traffic feature vectors for different malware categories. A proportion of the feature vectors extracted were used as training data to train few standard ML algorithms and the ML models thus obtained were applied to test data with their classification scores reported. As part of our future work, we are working on the software-based implementation of EDIMA and its performance evaluation. We are also planning to adapt some state-of-the-art botnet detection techniques using bot-CnC communication features and ML algorithms for malware activity detection and compare their performance with EDIMA.

\section*{Acknowledgment}
This research is supported by the National Research Foundation, Prime Minister’s Office, Singapore under its Corporate Laboratory@University Scheme, National University of Singapore, and Singapore Telecommunications Ltd.

\bibliographystyle{ieeetran}
\begingroup
\raggedright
\bibliography{oqeprop}
\endgroup

\end{document}